\documentclass[pra,aps,amsmath,twocolumn,showpacs,floatfix]{revtex4-1}
\usepackage{amsmath,graphicx}
\begin{document}
\def\tr{\rm{Tr}}
\def\la{{\langle}}
\def\ra{{\rangle}}
\def\a{{\alpha}}
\def\e{\epsilon}
\def\q{\quad}
\def\w{\tilde{W}}
\def\t{\tilde{t}}
\def\a{\hat{A}}
\def\h{\hat{H}}
\def\E{\mathcal{E}}
\def\p{\hat{P}}
\def\u{\hat{U}}
\def\n{\\ \nonumber}
\def\j{\hat{j}}
\def\alph{a}
\def\vc{\underline{c}}
\def\vf{\underline{f}}
\title{Asking photons where they have been in plain language }
\author {D. Sokolovski$^{a,b}$}
\affiliation{$^a$ Departmento de Qu\'imica-F\'isica, Universidad del Pa\' is Vasco, UPV/EHU, Leioa, Spain}
\affiliation{$^b$ IKERBASQUE, Basque Foundation for Science, E-48011 Bilbao, Spain}
\date{\today}
\begin{abstract}
The authors of  Phys.Rev.Lett.   \textbf{111}, 240402 (2013) conclude that "the past of the photons is not represented by continuous trajectories".
A simple analysis by standard quantum mechanics shows that this claim is false.
\end{abstract}
\maketitle
\noindent
{Keywords; {\it quantum particle's  past, Feynman paths, weak measurements}}
\section{Introduction}
Recently Danan and co-workers \cite{vPRL} 
reported what they consider an experimental confirmation of a surprising 
suggestion that 
photons do not always follow continuous trajectories, and can occasionally be found in the parts of a Mach-Zender interferometer (MZI) which they neither enter, nor leave. 
The suggestion itself does appear strange, if not wrong,  and has since been questioned by several authors.
It was disputed by L. Saldanha \cite{Sald} who presented a detailed analysis of photon's propagation in an MZI with rotating mirrors. Somewhat curiously, Saldanha described his analysis as
"complementary to the one using the two-state vector formalism (TSVF) of quantum theory" despite his apparent disagreement with the findings 
of  \cite{vPRL} where the formalism was indeed used. 
The TSVF approached fared less well with the authors of \cite{HIST0}, who found it perhaps not suitable for describing delicate interference effects such as  discussed here. 
A second-quantised version of Saldanha's argument was presented in \cite{HIST1}, where it was argued that 
the quantity used in \cite{vPRL} is not a 
reliable which-path witness. 
A simple illustration was given in \cite{HIST2}, where the authors showed that an MZI setup can be altered to provide the evidence of photons entering the previously inaccessible parts without changing the values of the said which-path witnesses.
Further criticism can be found  in \cite{Salih}, where it was pointed out that the relevant experimental evidence disappears, should the experiment be modified slightly. The authors of \cite{vPRL}
responded by noting that what matters is that the evidence can be present, not that it may be
  absent under certain conditions \cite{vCOMM}. 
  Finally, a critical analysis of a similar MZI setup in terms of the consistent histories approach can be found in a recent work by Griffiths \cite{Griff}.
 In the end, one may be left with a feeling that something fairly fundamental is being analysed at a perhaps too technical level, 
and that at least a part of the matter remains open for further discussion.
\newline
The original suggestion pointing to the photon's "surprising" behaviour was made in an earlier paper by Vaidman  \cite{v2013}, who used the 
"weak measurements formalism" of \cite{Ah} to investigate the photon's past. 
The "weak measurement reality" advocated by Vaidman in \cite{vREAL} has a simple explanation in standard quantum mechanics. A particular perturbative scheme realised by an inaccurate "weak" von Neumann meter \cite{vN} determines 
the values of the transition amplitudes (or their linear combinations)  for a pre- and post-selected system \cite{PLA2016}. Such amplitudes are well known to be  the basis of quantum description \cite{FeynL}. By adding them together, and squaring the modulus of the result, one finds the probabilities with which a certain sequence of observable events occurs should the experiment be repeated many times over.
The manner in which the amplitudes are added reflects what is being done to the system to destroy (or preserve) the interference between 
certain alternatives \cite{FeynL}. Seen from this prospective, the weak measurements approach adds little to what is 
already known from the textbooks (see, for example, \cite{FeynL}, \cite{Feyn}). 
\newline
Mere rebranding transition amplitudes as "weak values" would have no further consequence, had not some authors attributed to the 
weak values qualities transition amplitudes do not normally have. This led to a number of intriguing claims,  all allegedly supported by the weak measurements argument.  The claims included the evidence of
"negative kinetic energy" \cite{NKE}, "negative number of particles" \cite{AhHARDY}, "having one particle in several places simultaneously" \cite{AhBOOK}, "photon disembodied from its polarisation" \cite{CAT}, 
"electron with disembodied charge and mass" \cite{CAT}, and "an atom with the internal energy disembodied from the mass" \cite{CAT},
Earlier we argued \cite{PLA2016}, \cite{DS1}-\cite{MATH} that most of these notions are easily dismissed once the discussion is returned to the framework of standard quantum mechanics. 
We ask whether the same could be said about the strange behaviour attributed to the photons in \cite{vPRL} and  \cite{v2013}.
The purpose of this paper is, therefore, to see whether the assertions of \cite{vPRL} stand up to scrutiny when examined within the conventional quantum formalism.
\section{Standard quantum rules}
We start by recalling some of the quantum mechanical orthodoxy, as outlined in the opening chapters of the 
well known textbooks  \cite{FeynL} and  \cite{Feyn}.
At a given time an isolated quantum system is described by a state in its Hilbert space, $|\psi\ra$. We are, however, interested in a part of its history pertaining instead to a time interval, e.g., in a situation where the system starts in a state $|\psi\ra$ at $t=0$, and ends up in a state $|\phi\ra$ at $t=T$. Usually we can define several ways, or {\it paths}, via which the system can reach $|\phi\ra$, and for any choice quantum theory provides the {\it probability amplitudes}, $A[path]$. Let there be $N$ such paths, which we denote as $\{i\}$, $i=1,2..N$.
\newline
A paths is {\it real}, if also a probability $P[path]$, usually given by the Born rule $P[path]=|A[path]|^2$, can be ascribed to it. If an experiment is repeated many times, the system will be seen "travelling" the path with a relative frequency proportional to $P[path]$.
\newline
A path is {\it virtual}, if we can ascribe to it only an amplitude $A[path]$, but not a probability $P[path]$. Such a path may enter, together with other virtual path(s), as a constituent part of a composite real pathway, e.g., $\{1+2\}=\{1\}+\{2\}$, travelled with the probability $P[1+2]=|A[1]+A[2]|^2$. The manner in which an individual path amplitudes are added before squaring the module of the result indicates that the paths are {\it interfering} rather than {\it exclusive} alternatives \cite{Feyn}. The rule can be extended to the paths constructed as weighted combinations (superpositions) of other paths. For example, a path $$\{\alpha 1+ \beta 2\}\equiv \alpha \{1\}+ \beta \{2\}$$ carries an amplitude
$$A[\alpha 1+ \beta 2]=\alpha A[1]+\beta A[2],$$ where $\alpha$ and $\beta$ are two complex numbers.
\newline Feynman's uncertainty principle (UP) \cite{FeynL} implies that two, or more, interfering paths cannot be distinguished without destroying the interference between them. They form, in a strong sense, a single real route, and together should be taken "as a single unit" \cite{Kast}.
\newline
Suppose that all $N$ paths in our example are virtual and, according to the UP, we only know that the system reached in $|\phi\ra$, but cannot say how. We may enquire about the system's past by coupling it to a probe, e.g., a meter which would determine the system's state at $t=T/2$. If the measurement can provide $M\le N$ different outcomes $m_i$, $i=1,2,...,M$, we will have created $M$ real routes connecting $|\psi\ra$, $|\phi\ra$, $|\phi\ra \gets m_i \gets |\psi\ra$. Adding more meters at different times, will create a stochastic network of real paths, travelled by the system with relative frequencies $w(i)$, evaluated as discussed above \cite{MATH}. The emphasis here is on the word "create". The real paths were not there before, and have been "constructed" by the meter(s), using the available virtual paths of the system as building elements. 
\newline
Assigning a number (a functional) $F[i]$ to each real path, we can calculate its {\it mean value}, $\la F\ra=\sum_{i=1}w(i)F[i]$, 
and measure it by writing down the value $F[i]$ whenever the meter reads $m_i$, summing the results, and dividing them by the 
number of times the experiment is performed.
\newline
Let the meter be a von Neumann pointer with position $f$ \cite{vN}, prepared in a state $G(f/\Delta f)$, where real $G(f)$ has is bell-shaped, peaking at $f=0$, and drops of rapidly as $|f|\to \infty$. Suppose also that we want to measure an operator (a projector), taking the value $1$ on the path number $1$, $F[1]=1$, and $0$ on all others, $F[i\ne1]=0$. Now the probability to find a pointer reading $f$ is given by \cite{PLA2016}
\begin{eqnarray}\label{1}
P(f)=\frac{\rho(f)}{\int \rho(f)df}, \q \rho(f)\equiv \left |\sum_{i=1}^N G\left (\frac{f-F[i]}{\Delta f}\right )A[i]\right |^2\q\q
\end{eqnarray} 
\newline 
Noting finally that the paths leading to distinguishable final states, e.g., to $|\phi_1\ra$ and $|\phi_2\ra$ such that $\la \phi_1|\phi_2\ra=0$, cannot interfere and are, therefore, exclusive alternatives, we have almost all the necessary pre-requisite for analysing the photon's past in the experiments proposed in \cite{v2013}, and performed in \cite{vPRL}. 
\section{Accurate vs. inaccurate (Strong vs. weak)}
What we will learn about the system's past crucially depends on the accuracy $\Delta f$, with which  the pointer has been set to zero.
For $\Delta f<<|F[1]-F[2]|=1$ the initial uncertainty is small, the measurement is accurate ({\it strong}), and the meter creates two real paths, $\{I\}=\{1\}$, and the superposition of all the remaining ones, $\{II\}=\{2+3+...+ N\}$, with amplitudes
\begin{eqnarray}\label{2}
A[I]=A[1],\q and \q A[II]=A[2]+...+A[N].
\end{eqnarray} 
The paths are travelled with the relative frequencies $w(I)=|A[I]|^2/(|A[I]|^2+|A[II]|^2)$ and
$w(II)=|A[II]|^2/(|A[I]|^2+|A[II]|^2)$, respectively. The mean value of the projector is just the relative frequency with which the (now real) first path is travelled, $\la F\ra=w(I)$.
\newline
Making the meter highly inaccurate by sending $\Delta f \to \infty$, helps to minimise the perturbation incurred on the system, leaving the probability to go from $|\psi\ra$ to $|\phi\ra$ almost untouched, whatever the meter's reading $f$,
\begin{eqnarray}\label{3}
\rho(f)= G\left (\frac{f}{\Delta f}\right )^2|A[I] + A[II]|^2+O(1/\Delta f).
\end{eqnarray} 
Now the possible  meter's readings fill the whole of the real axis, $-\infty \le f \le \infty$, rather than being $1$ or $0$, 
as in the accurate strong case. This gives an operational meaning to the uncertainty principle \cite{MATH}. Asking a meaningless question "which way was taken, with the interference intact?", provokes an answer "anything at all".
\newline
 Following \cite{Ah} we note, however, that
the mean meter reading, $\overline f \equiv \int f P(f) df$, tends to a finite limit as $\Delta f \to \infty$. For our projector onto the  first path,  this limit is \cite{PLA2016},\cite{MATH} 
\begin{eqnarray}\label{4}
\lim_{\Delta f \to \infty}\overline f=Re  \alpha[I], \q \alpha[I]\equiv \frac{A[I]}{A[I]+A[II]}.
\end{eqnarray}
The average can, in principle, be evaluated after many trials \cite{Ah}. It is, therefore, possible to extract the values of a {\it relative path amplitude}, $\alpha[I]$.
Thus, an inaccurate ({\it weak}) measurement of a projector determines the probability amplitude for the path, for which an accurate measurement finds the frequency with which the path is travelled \cite{PLA2016}. There is no contradiction with the Uncertainty principle:
knowing the value of an amplitude allows one to predict what {\it would }happen, should a virtual path be made real by interaction with an external agent.  It by no means implies that this interaction {\it has} taken place. If necessary, more detailed discussion of the above principles can be found, for example, in \cite{MATH}.
\section{Where was the photon? (The weak trace argument)}
We can now give an answer to this question, as posed by Vaidman  in \cite{v2013}. Consider a system of nested Mach-Zehnder 
interferometers (MZI) or, rather, its part connecting the input with the output at the detector D. 
\begin{figure}
	\centering
		\includegraphics[width=8.0cm,height=8.0cm]{{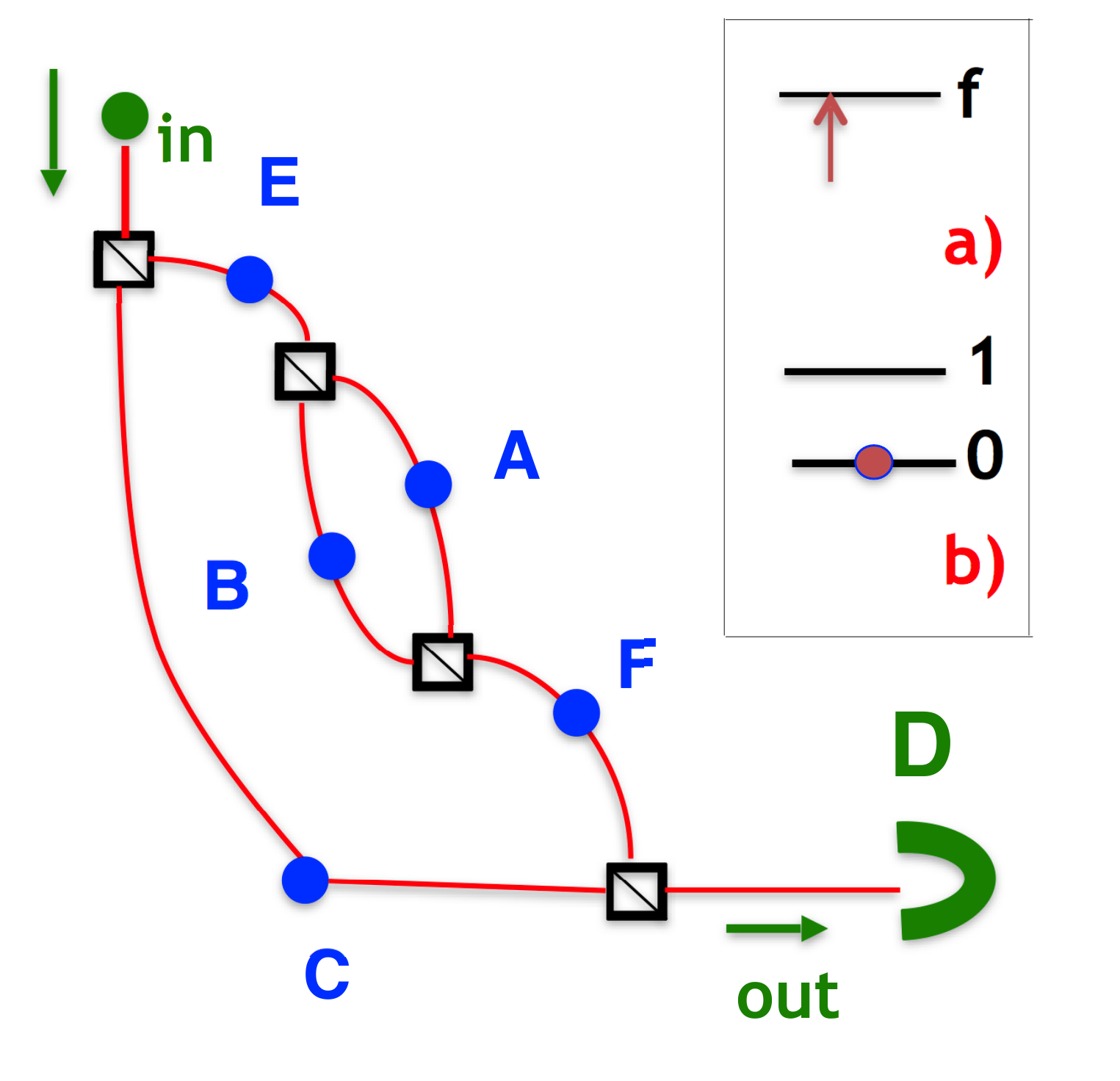}}
\caption{(Color online) 
Part of a system of nested Mach-Zehnder interferometers leading to the detection  of an injected photon by a detector $D$.
After two reflections in the two inner beam splitters, a photon passing through the branch $A$ enters the branch $F$ with an additional 
phase of $\pi$.
Two types of measurements made at the points $A$, $B$,$C$, $E$, and $F$ are illustrated in the inset:
a) by a weak von Neumann pointer (arrow) with a position $f$, and b) by a two-level system, capable of changing its state
as a photon passes by.}
\label{fig:1b}
\end{figure}
A photon enters in a wave packet state $|in\ra=|\psi\ra$, and travels each arm of the MZI suffering neither delays no distortions (or  undergoing the same delays and distortions in all three) to arrive at D in a state
\begin{eqnarray}\label{a}
|out\ra =|\phi\ra\{A[1]+A[2]+A[3]\} 
\end{eqnarray}
where $A[i]$, are the amplitudes for the three virtual paths $\{i\}$, $i=1,2,3$ (see Fig.2), 
\begin{eqnarray}\label{5}
\{1\}: \q by\q going \q D \gets F\gets A\gets E \gets in,\n
\{2\}: \q by\q going \q D \gets F\gets B\gets E \gets in,\n
\{3\}:  \q by\q going  \q D \gets C \gets in,\q\q\q\q\q
\end{eqnarray}
leading to $D$. Detector $D$ clicks whenever it finds the photon (post-selected) in the state $|\phi\ra$.
\newline
Let weak von Neumann pointers, $M_A$, $M_B$,$M_C$, $M_E$, and $M_F$,  (see the inset in Fig.1) be locally coupled to the photon
at the points $A$,$B$,$C$,$E$, and $F$ shown in Fig.1.
Each pointer may be shifted whenever the photon passes through the corresponding point in its arm of the interferometer,
and remains in its place otherwise. 
Provided the MZI are tuned so that 
\begin{eqnarray}\label{6}
A[1]=-A[2]\ne 0,  \q and \q A[3]\ne 0,
\end{eqnarray}
and  given that the detector $D$ clicks, which of the pointers $f_i$, $i=A,B,C,E,F$, would register a non-zero 
average reading, $\overline{f}_i\ne 0$, if many trials are made? Or, in the words of Ref.  \cite{v2013}, where would a passing photon leave a {\it weak trace}?
\newline
Conventional quantum mechanics offers a straightforward answer. Since our weak pointer measures the  amplitudes where a strong one measures the frequencies, the $i$-th  weak trace will be left wherever a strong measurement made by $M_i$, with all other meters disconnected, would give a non-zero result. Indeed, for a strong pointer to move we need the amplitude on the path in which it detects the photon to be non-zero. But then the average reading must also be non-zero by (\ref{4}). (We could have added "and $Re A[i] \ne 0$, but a slightly different setup allows to measure, in a similar way,  $Im \alpha_i$ \cite{Ah},\cite{PLA2016}, so all we need is that an $A[i]$ doesn't vanish.) 
\begin{figure}
	\centering
		\includegraphics[width=4.5cm,height=6.5cm]{{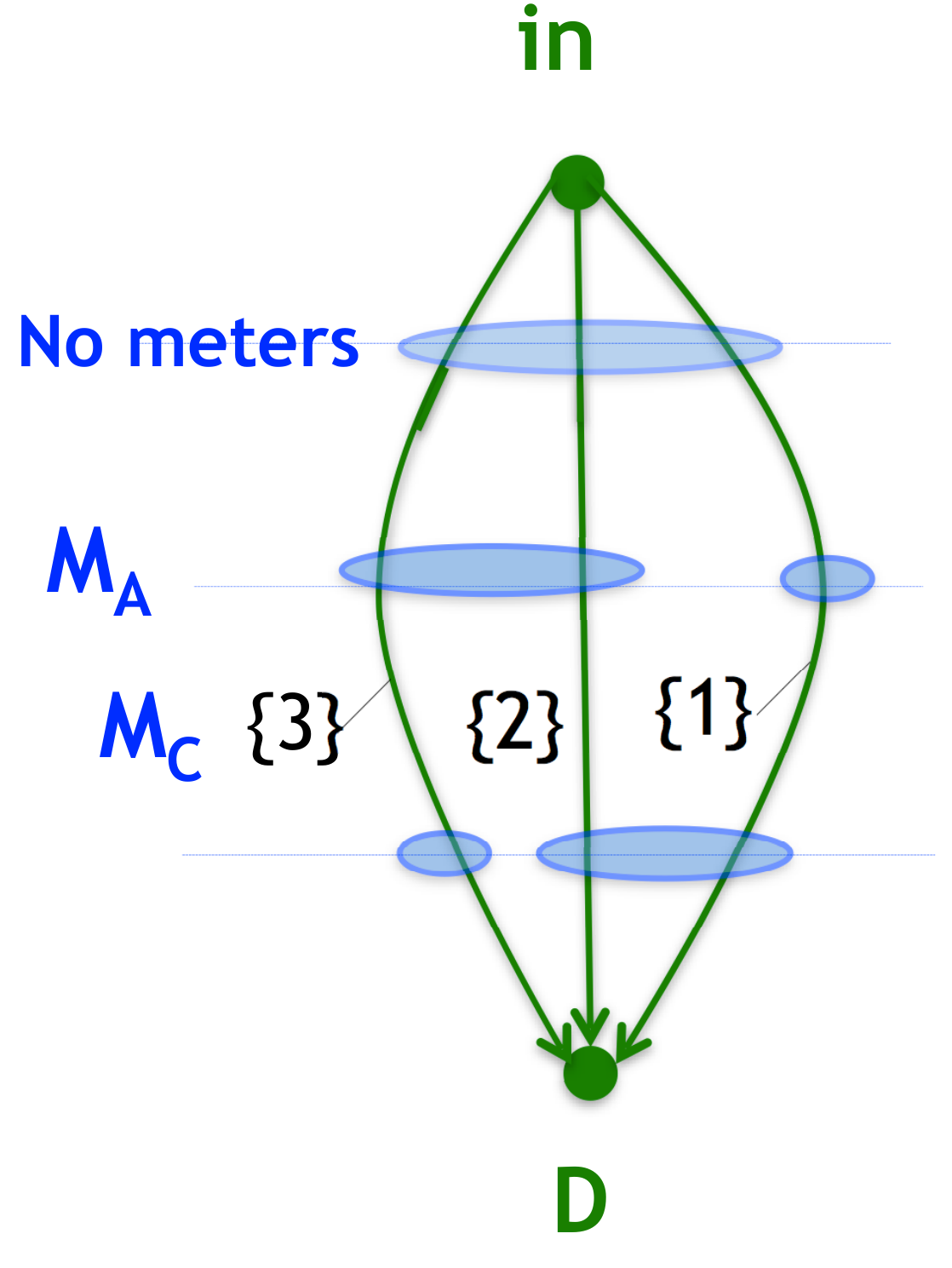}}
\caption{(Color online) 
For a photon, entering the setup in Fig.1, the three virtual paths in (\ref{5}) form a single pathway, $\{1+2+3 \}$ leading to 
detection in $D$. A measurement by $M_A$ combines the paths $\{2\}$ and $\{3\}$ into a pathway $\{2+3\}$.
Similarly, for $M_C$ there are two pathways, $\{I\}=\{3\}$ and  $\{II\}=\{2+3\}$.} 
\label{fig:2b}
\end{figure}
For example, the meter $M_A$ creates two real paths, $\{I\}=\{1\}$ and $\{II\}=\{2\}+\{3\}$, and detects the photon in $\{I\}$, $F[I]=1$, $F[II]=0$. Since
$A[I] =A[1]\ne 0$, a weak trace well be left in the arm labelled $A$. By the same token, a trace will also be left in the arm $B$.
\newline The meter $M_C$ detects in the path $\{I\}=\{3\}$, with $A[3]\ne 0$, hence there will be a weak trace in the arm $C$.
\newline Finally, the meters $M_E$ and $M_F$ both detect in $\{I\}=\{1\}+\{2\}$, with $A[I]=A[1]+A[2]=0$, so there will be no weak trace left in either $E$ or $F$.
\newline
We note that our simple result agrees with the predictions of \cite{v2013}. Next we consider a slightly different setup, similar to the one used in \cite{vPRL}, in order to see if we recover  the "surprising" results reported in \cite{vPRL}.
\section{Where was the photon? (By leaving marks) }
In  \cite{v2013} Vaidman proposed an alternative scheme, in which a photon "leaves a mark" at a particular place by 
changing the internal state of a two-level system, locally coupled to it there. Next we consider such a scheme in some detail.
In the setup shown in Fig.1 we replace the weak pointers at locations $A$,$B$, ..., $F$ with two-level systems, initially in the ground states $|0_X\ra$, $X=A,B...$,  [see Fig.1 inset b)]. A passing photon may excite the system  to $|1_X\ra$  (a "mark" is left),with an amplitude $a_1^X$,  or leave it in the ground state $|0_X\ra$ (no mark left), with an amplitude $a_0^X$. We want $|a_1^X|$ to be small, so as to perturb the transition only slightly (see Appendix A for details).
(This may not be a good practical way to describe a photon's interaction, but will serve for our {\it gedankenexperiment}
no worse than the hypothetical weak von Neumann pointer of \cite{v2013} and the previous Section.)
With the reflection negligible, the interaction with a photon entering in a state $|\phi\ra$ results in an entangled state, 
\begin{eqnarray}\label{g1}
|\phi\ra|0_X\ra \to |\phi\ra(a^X_0|0_X\ra+a^X_1|1_X\ra), 
\end{eqnarray}
where $|a_0^X|^2+|a_1^X|^2\approx1$. Following \cite{vPRL}, we inject a large number of photons, select only the cases in which the detector $D$ clicks, and look for the marks that have been left, indicating the presence of a photon at that point.

There are now $2^5=32$ orthogonal final states, $|\phi\ra|i_A\ra|i_B\ra|i_C\ra|i_E\ra|i_F\ra\equiv |\phi,i_A,i_B,i_C,i_E,i_F\ra$, 
$i_X=0,1$ and, in principle, $32$ real pathways leading to them, as shown in Fig. 3. The paths are travelled with the probabilities 
$P(i_A,i_B,i_C,i_E,i_F)=|A(i_A,i_B,i_C,i_E,i_F)|^2$, i.e., the probabilities to have a click, and find 
the system at $X$  in  a state $|i_X\ra$, $i_X=0,1$.
Using (\ref{g1}), and taking into account what may happen along each of the the three paths shown in Fig.2, 
we find the final state of the photon to be
\begin{eqnarray}\label{g2}
|out\ra =|\phi\ra\{A[1](\alph^E_0|0_E\ra+\alph^E_1|1_E\ra)
\q\q\q
\q\q\q\q\q\q\q\n
\times (\alph^A_0|0_A\ra+\alph^A_1|1_A\ra)(\alph^F_0|0_F\ra+\alph^F_1|1_F\ra)|0_B\ra|0_C\ra
\q\q\q\q\q\n
+A[2](\alph^E_0|0_E\ra+\alph^E_1|1_E\ra)\q\q\q\q\q\q\q\q\q\q\q\q\q\q\q\n
\times(\alph^B_0|0_B\ra+\alph^B_1|1_B\ra)(\alph^F_0|0_F\ra+\alph^F_1|1_F\ra)|0_A\ra|0_C\ra\q\q\q\q\q\n
+A[3](\alph^C_0|0\ra+\alph^C_1|1\ra)|0_E\ra|0_A\ra|0_B\ra|0_F\ra\}.\q\q\q\q\q\q\q\q\q
\end{eqnarray}
\begin{figure}
	\centering
		\includegraphics[width=3.5cm,height=5.5cm]{{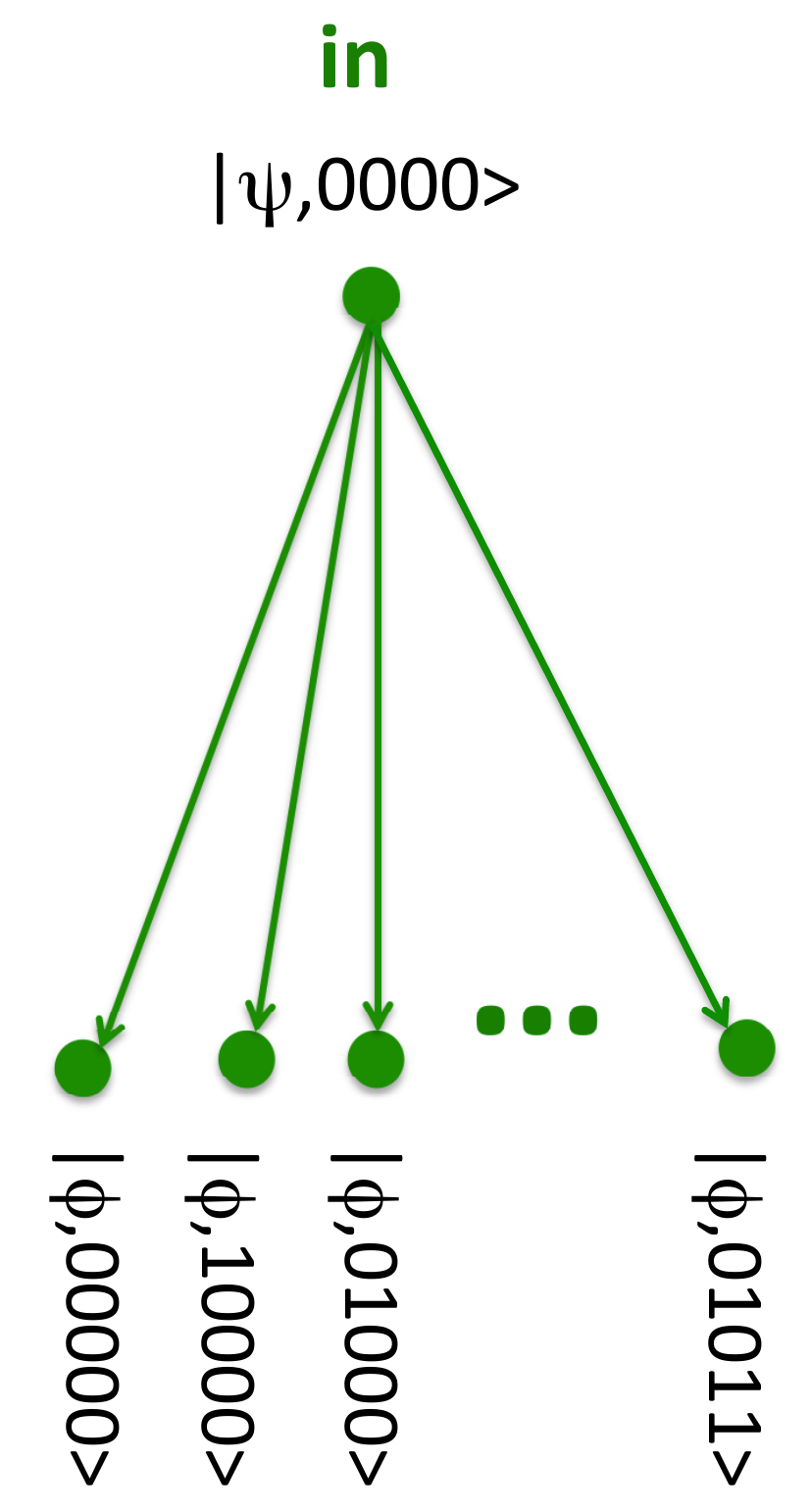}}
\caption{(Color online) Real pathways connecting the initial state with $13$ distinguishable outcomes.
Ones and zeroes indicate the states of the two-level markers, written in the order $A$,$B$, ...., $F$. } 
\label{fig:3b}
\end{figure}
The rest is a straightforward exercise in book keeping. The way in which the two-level markers are placed reduces the number of relevant paths from $32$ to $13$, since some of the final states just cannot be reached.
[For example in (\ref{g2}), there is no term containing a product $|1_A\ra|1_B\ra$, so that a photon cannot be detected both in the arms $A$ and $B$ and, similarly, in $A$ and $C$, or $B$ and $C$.]
Among the remaining possibilities, there is one to avoid leaving any marks at all. There are three
ways to do so.
Firstly, by the path $\{3\}$, avoiding leaving a mark at  $C$, with an amplitude $A[3]a_0^C$. Secondly, via the path $\{1\}$, leaving no marks at $E$,
$A$ and $F$, with an amplitude $A[1]a_0^Ea_0^Aa_0^F$. Thirdly, via the path $\{2\}$, leaving no marks at  $E$,
$B$ and $F$, with an amplitude $A[2]a_0^Ea_0^Ba_0^F$. Adding the three, we obtain the probability amplitude
\begin{eqnarray}\label{g3}
A(0,0,0,0,0)=A[1]a_0^Ea_0^Aa_0^F+A[2]a_0^Ea_0^Ba_0^F+A[3]a_0^C, \q\q
\end{eqnarray}
for the real pathway $a_0^Ea_0^Aa_0^F\{1\}+a_0^Ea_0^Ba_0^F\{2\}+a_0^C\{3\}$. Here the arm of the MZI chosen by the photon 
remains indeterminate, just as the slit chosen by an electron in the Young's double-slit experiment \cite{Feyn}.
\newline 
Similarly, there are two ways to leave  a mark at $F$, but not anywhere else. The amplitudes for leaving no marks at   $E$ and $A$, and  $E$ and $B$ on the paths $\{1\}$ and $\{2\}$, are $A[1]a_0^Ea_0^Aa_1^F$ and $A[2]a_0^Ea_0^Ba_1^F$, respectively.
The probability to leave a mark only at $F$ is, therefore, 
\begin{eqnarray}\label{g4}
P(0,0,0,0,1)=|A[1]a_0^Ea_0^Aa_1^F+A[2]a_0^Ea_0^Ba_1^F|^2, \q\q
\end{eqnarray}
and the uncertainty principle prevents us from knowing whether the arm $A$ or the arm $B$ has actually been travelled.
In a similar way, we are able to evaluate probabilities for all $13$ paths, listed in the Appendix B.

Now we want to repeat the experiment may times, and to see where marks will be left, and how often.
The net probability to find a mark at, say, $A$ is given by
\begin{eqnarray}\label{g5}
W(A) = \sum_{i_B,i_C,i_E,i_F=0}^1P(1,i_B,i_C,i_E,i_F), \q\q
\end{eqnarray}
and similarly for $W(B)$,...,$W(F)$.
\section{So where has the photon been? }
Next we want to learn something about photon's past from analysing the probabilities $W(X)$ in Eq.(\ref{g5}). We choose \cite{FOOT}
\begin{eqnarray}\label{g6}
A[1]=-A[2]=\sqrt{1/12}, \q
A[3]= \sqrt{1/6},
\end{eqnarray} 
 and 
 \begin{eqnarray}\label{g7}
a_1^X=-0.05i\equiv -i\epsilon, \q X=A,...,F.
\end{eqnarray}

The results (smeared with a Gaussian, admittedly to make comparison with Fig.2 of \cite{vPRL} easier) are shown in Fig. 4.
At first glance they seem to support the conclusions made by the authors of \cite{vPRL}. The marks are seen to have been left 
at $A$, $B$, and $C$, but there appears to be no trace of the photons at either $E$ or $F$. A similar argument led the authors of \cite{vPRL} to insist that { \it "The photons do not always follow continuous trajectories. Some of them have been inside the nested
interferometer...., but they never entered and never left
the nested interferometer."}

Standard quantum mechanics gives, however, a different account. The photon passes through one of the exclusive pathways, 
real and equipped with probabilities, which were
 created by the interaction with the two-level systems installed to monitor its progress. 
Not all of the paths coincide with one of the arms of the interferometer, as it happens, for example, in the case where no marks are left [cf. Eq.(\ref{g3})]. Strange as it may seen to a classical observer, these pathways are {\it real} to a quantum photon, which chooses one of them with a frequency proportional to $P(i_A,i_B,i_C,i_E,i_F)$.
\newline
Thus, there is a fault with our first-glance assumption.
A closer look shows that the probabilities $W(E)$ and $W(F)$ are not zero, but are
 extremely small (see the inset in Fig.4). 
We can easily find out where these small peaks come from by counting the powers of the small parameter $\epsilon$ in Eq.(\ref{g7}).
Since  the amplitudes on the last three
paths in the list in the Appendix B vanish by (\ref{g6}) and  (\ref{g7}), the probabilities of the scenarios where a mark is left 
at only $E$ or $F$ (paths numbered   $4$, $5$, $8$, and $9$ in the Appendix B) are of oder of $\epsilon^4$.
The chances to leave a mark at both $E$ and $F$ (paths $6$ and $10$) are even slimmer, $\sim \epsilon^6$.
and can be easily overlooked.
The picture of a photon circling the loop of the embedded MZI without 
entering or leaving it \cite{vPRL} arises, therefore, only in the leading-order approximation, 
beyond which the authors of \cite{vPRL} were discouraged to look by the misleading analysis of \cite{v2013}.
It  doesn't stand up to scrutiny if 
an accurate inspection of the relevant paths is made.
\begin{figure}
	\centering
		\includegraphics[width=8.5cm,height=6cm]{{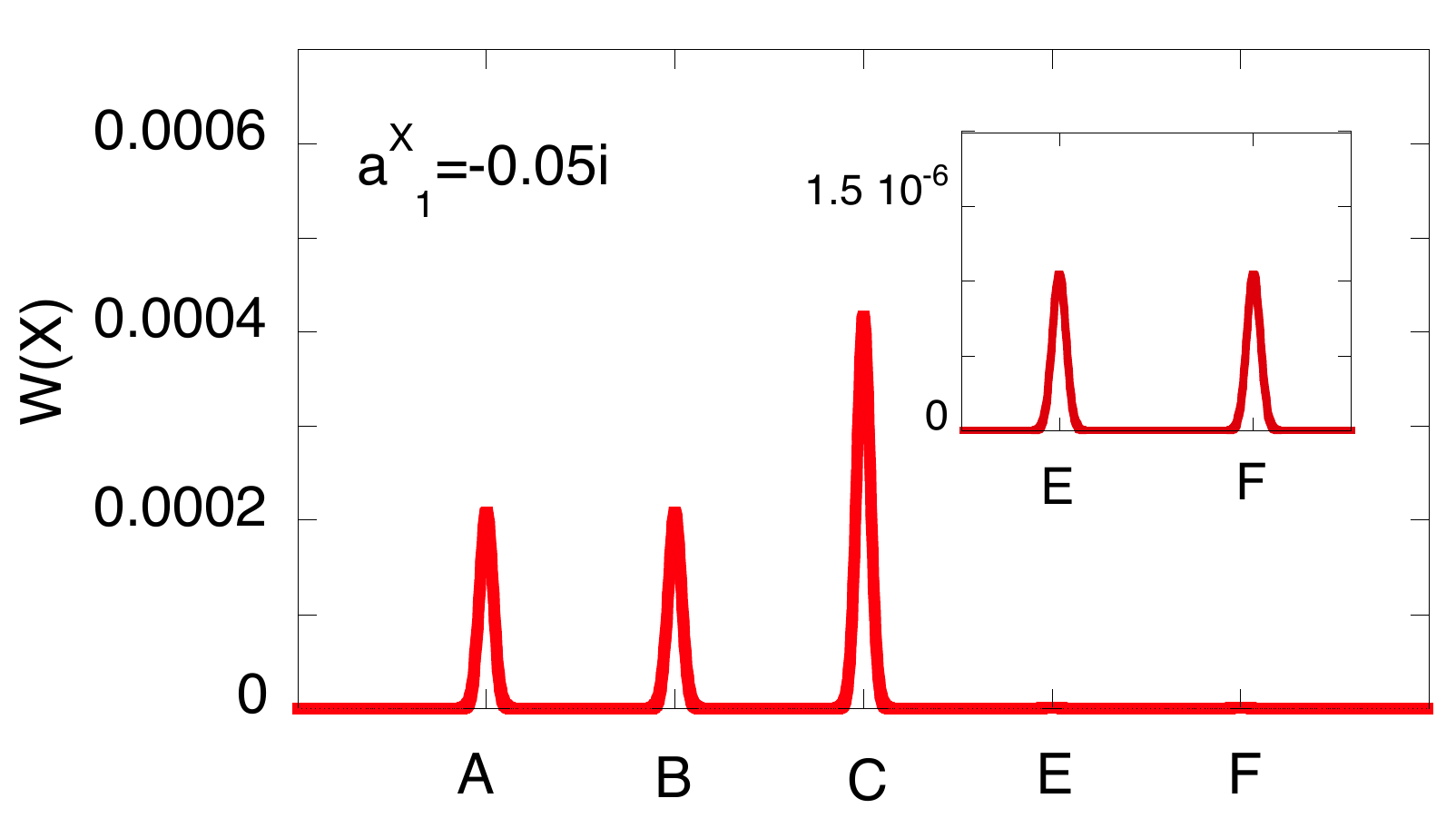}}
\caption{(Color online) 
The probabilities $W(A)$, $W(B)$,...,$W(F)$in Eq.(\ref{g5}). In the inset are shown $W(E)$ and $W(F)$, too small to be visible
in the original plot.} 
\label{fig:4b}
\end{figure}

\section{One more time, where did the photon go? (By making a change, and seeing where it matters) }
Our discussion would be incomplete if we left out the analysis of Saldanha \cite{Sald} of the experiment described in \cite{vPRL}.
The analysis does not involve external degrees of freedom. Rather, the markers at $A$, ...,$F$ are replaced by mirrors, which can 
be rotated. The question is: which of the small rotation would affect the signal at the detector $D$?
To avoid going into the technical details, which can be found in \cite{Sald}, we ask a simpler question.
To follow the path $\{1\}$ in Fig.2, the photon has to travel the arm $E$, and then the arm $A$, followed by the arm $F$.
Let $A[X]$, $X=E,A,F$ be the corresponding amplitudes. Then, by the rule of multiplication, the amplitude to travel the path $\{1\}$
is $A[1]=A[E]\times A[A]\times A[F]$, and similarly for the paths $\{2\}$ and $\{3\}$. Let us introduce localised perturbations, such that each amplitude $A[X]$ changes a little, $A'[X]=A[X]+\delta[X]$, and see how this would affect the probability $P$ to arrive in the detector $D$.
We now have 
\begin{eqnarray}\label{k1}
A'[1]=(A[E]+\delta[E])(A[A]+\delta[A])(A[F]+\delta[F])\q\n
A'[2]=(A[E]+\delta[E])(A[B]+\delta[B])(A[F]+\delta[F]),\n
A'[3]=(A[C]+\delta[C]),
\end{eqnarray}
and $P=|A'[1]+A'[2]+A'[3]|^2$. The photon going through the arm $A$ is twice reflected at the internal beam splitters, each time acquiring 
a phase of $\pi/2$, so with the appear part of the MZI blocked, $A[1]=-A[2]$, we should also have $A[A]=-A[B]$. The perturbations $\delta[X]\sim \epsilon$ are chosen to be small so as not to perturb the transition, and to see how each each perturbation affects the $P$ we will count the powers of $\epsilon$, in the same way we did in the previous Section. In the first order approximation we obtain
(cf. Eq.(7) of \cite{Sald})
\begin{eqnarray}\label{k2}
A'[1]+A'[2]+A'[3]=A[3]+\q\q\q\n
A[E](\delta[A]+\delta[B])A[F] +\delta[C] +o(\epsilon).
\end{eqnarray}
Notably, $\delta[E]$ and $\delta[F]$ are absent from the r.h.s of Eq.(\ref{k2}) and, encouraged by the predictions of \cite{v2013}, we may be tempted to conclude that the photon has not been in either $E$ or $F$ (perturbations introduced there have no effect), 
while it has been in $A$ and $B$ (perturbations there do have an effect). That would be a mistake. Equation (\ref{k2}) is only approximate, and the evidence of the photon visiting both $E$ and $F$ appears already if the terms quadratic in $\epsilon$ are kept.
In particular adding to the r.h.s. of (\ref{k2}) the remaining terms we have
\begin{eqnarray}\label{k3}
A[E](\delta[A])+\delta[B])\delta[F]+\delta[E](\delta[A]+\delta[B])A[F].\q\q\q\n
+\delta[E](\delta[A])+\delta[B])\delta[F],
\end{eqnarray}
and there is no reason why the quantity in Eq.(\ref{k3}) should vanish, or make a vanishing contribution to the probability $P$. (Choosing $\delta[A]=-\delta[B]$ would make it vanish but with, in Eq.(\ref{k2}),  it would disappear also the evidence of the photon having been in the arms $A$ and $B$).
Thus, the conclusion is similar to that of the previous Section. The evidence of the photons passing through the arms $E$ and $F$
may be harder to find, but it is there in general. 

\section{Conclusions and discussion }
From the conventional point of view, an observed quantum particle has been in one of the exclusive pathways, made available to it by the interaction with the measuring devices. 
For each such scenario, quantum theory provides an amplitude, and also a probability. 
This probability determines the frequency with which a scenario occurs, should the experiment be repeated many times.
Thus, our conclusion is similar to that of the previous Section. 

If inaccurate measurements of Sect. III are performed, and the limit in Eq.(\ref{4}) is duly taken, 
the transition is perturbed only weakly. The particle remains in a real pathway combining all interfering paths, 
and the pointer readings, averaged over many trials, coincide with the real or imaginary parts of the corresponding probability amplitude. This is not unexpected, since one can often deduce both the moduli and phases of transition
amplitudes from the response of quantum observables to a small external perturbations \cite{PLA2016}.
 It is wrong, however, to assume that the particle has been in one of the virtual paths and not in the other.
 Knowing an amplitude allows one to predict what {\it would} happen if an accurate meter is employed, 
 but by no means implies that someone has indeed brought this additional piece of equipment into the lab, and turned it on.
For this reason, the weak measurement argument cannot support, for example, the claim that the photons {\it " ...have been inside the nested interferometer (otherwise they could not have known the frequencies $f_A$, $f_B$), but they never entered and never left the nested interferometer...."} \cite{vPRL}.
  or {\it "...do not always follow continuous paths,..."} \cite{v2013}.
 The calculation in \cite{v2013} is correct, but the conclusion drawn from it is unjustified.

The case of two-level markers employed to monitor the photon's progress can be analysed in a similar way.
The possibility to change the internal state of a marker creates for the system a variety of real pathways, some coinciding 
with the arms of the MZI, and some not. As the interaction becomes smaller, the most frequent scenario is where interference between different arms is left intact, and the "which arm?" question remains unanswered 
in accordance with the uncertainty principle. Among other much less probable outcomes are those in which the photon actually travels through the arm $F\gets A\gets E$, $F\gets B\gets E$, or $C$. There are also paths which are  superpositions of $F\gets A\gets E$ and $F\gets B\gets E$. The special choice $A[1]=-A[2]$ and $a_1^A=a_1^B$
in Eqs.(\ref{g6}) and Eqs.(\ref{g7})
serves to block such paths, and makes the chance of leaving a mark at $E$ or $F$ extremely small (see Fig.4).
Crucially, this probability cannot be made exactly zero, and the evidence of photons entering and leaving the inner MZI
is always present, contrary to the claims made in  \cite{vPRL}, \cite{vCOMM},  and \cite{v2013}.

The same conclusion cab be drawn from the elementary perturbation theory of Sect. VII.
While our models are not identical to the experiment described in \cite{vPRL}, it might be a prudent advice for its authors  to double check
that  in Fig.2 of \cite{vPRL} the signal at the frequencies $f_E$ and $f_F$ is indeed absent, and not just too small to be seen against the background noise.

We summarise.
Firstly, the argument of \cite{v2013} doesn't justify the claim made about the discontinuous behaviour of the photons in the nested MZI setup.
Secondly, simple models, similar to the one used in \cite{vPRL}, do provide non-vanishing evidence of the photons entering and leaving the inner loop.
It is, therefore, highly likely that the picture describing the "surprising" behaviour of the photons   \cite{vPRL}, \cite{vCOMM},  and \cite{v2013} is a yet another 
artefact \cite{PLA2016} of resorting to exotic interpretations of quantum mechanics, without consulting first  with the standard quantum theory.
\section {Acknowledgements}  Support of
MINECO and the European Regional Development Fund FEDER, through the grant
FIS2015-67161-P (MINECO/FEDER)
I am also grateful to Lev Vaidman for bringing to my attention the still outstanding task of explaining the results of 
Ref. \cite{vPRL} within the framework of conventional quantum mechanics.
 \section{Appendix A}
 Consider a spin-1/2 coupled to a passing particle of a mass $1$ via $V_{int}=\Omega \sigma_x \delta(x-x_0)$, where
 $\sigma_x$ is the Pauli matrix, $\sigma_{x11}=\sigma_{x22}=0$, $\sigma_{x12}=\sigma_{x21}=1$, and $\delta(x)$ is the 
 Dirac delta. The particle with a momentum $k$ impacts from the left the spin, polarised in the $z$-direction, and causes it to flip. 
Writing the $z$-polarised state of the spin, $|+\ra_z$,  in terms of the states polarised along the $x$-axis, 
$|+\ra_z=(|+\ra_x+|-\ra_x)/2$, we note that if the spin is in the state $|\pm\ra_x$, the particle experiences at $x=x_0$  a zero-range potential 
$\pm \Omega \delta (x-x_0)$. The transmission amplitude for such a potential is $k/(k+i\Omega)$, and we easily find the  amplitudes $a_0$ and $a_1$ in the limit $\Omega\to 0$ to be
 \begin{eqnarray}\label{apb1}
a_0=k^2/(k^2+\Omega^2) \approx 1-\Omega^2/k^2\n
a_1= -ik\Omega /(k^2+\Omega^2)\approx -i\Omega/k.
\end{eqnarray}
The overall probability to be reflected is, therefore, $\Omega^2/(k^2+\Omega^2)$, which vanishes as $\approx \Omega^2/k^2$ 
for small $\Omega$'s. For a broad wave packet with a mean momentum $k_0$ we may replace $k$ with $k_0$ in the expressions for the amplitudes for flipping the spin and leaving it in its initial state.
 \section{Appendix B}
Here we give the list of the $13$ real paths in Fig.3 produced by weakly coupling two-level markers  at 
 the points $A$,$B$,$C$,$E$,$F$.
 \begin{eqnarray}\label{a1}
 \nonumber
1)\q\text{to} \q|\phi,00000\ra \q \text{via}\q
a_0^Ea_0^Aa_0^F\{1\}+a_0^Ea_0^Ba_0^F\{2\}\q\q\q\q\q\q\n
+a_0^C\{3\},\q\q\q\q\q\q\q\q\n
2)\q\text{to}\q|\phi,001000\ra \q \text{via}\q a_1^C\{3\},\q\q\q\q\q\q\q\q\q\q\q\q\q\\
3)\q\text{to}\q|\phi,10000\}\ra \q \text{via}\q a_0^Ea_1^Aa_0^F\{1\},\q\q\q\q\q\q\q\q\q\q\q\n
4)\q\text{to}\q|\phi,10010\ra \q\text{via}\q a_1^Ea_1^Aa_0^F\{1\},\q\q\q\q\q\q\q\q\q\q\q\n
5)\q\text{to}\q|\phi,10001\ra \q \text{via}\q a_0^Ea_1^Aa_1^F\{1\},\q\q\q\q\q\q\q\q\q\q\q\n
6)\q\text{to}\q|\phi,10011\ra \q \text{via}\q a_1^Ea_1^Aa_1^F\{1\},\q\q\q\q\q\q\q\q\q\q\q\n
7)\q\text{to}\q|\phi,01000\ra \q \text{via}\q a_0^Ea_1^Ba_0^F\{2\},\q\q\q\q\q\q\q\q\q\q\q\n
8)\q\text{to}\q|\phi,01010\ra \q \text{via}\q a_1^Ea_1^Ba_0^F\{2\},\q\q\q\q\q\q\q\q\q\q\q\n
9)\q\text{to}\q|\phi,01001\ra \q \text{via}\q a_0^Ea_1^Ba_1^F\{2\},\q\q\q\q\q\q\q\q\q\q\q\n
10)\q\text{to}\q|\phi,01011\ra \q \text{via}\q a_1^Ea_1^Ba_1^F\{2\},\q\q\q\q\q\q\q\q\q\q\q\n
11)\q\text{to}\q|\phi,00010\ra \q \text{via}\q a_1^Ea_0^Aa_0^F\{1\}+a_1^Ea_0^Ba_0^F\{2\},\q\q\q\q\q\n
12)\q\text{to}\q|\phi,00001\ra \q \text{via}\q a_0^Ea_0^Aa_1^F\{1\}+a_0^Ea_0^Ba_1^F\{2\},\q\q\q\q\q\n
13)\q\text{to}\q|\phi,00011\ra \q \text{via}\q a_1^Ea_0^Aa_1^F\{1\}+a_1^Ea_0^Ba_1^F\{2\}.\q\q\q\q\q
\end{eqnarray}
The path amplitudes are easily calculated by replacing in the r.h.s. of (\ref{a1}) a path $\{i\}$, $i=1,2,3$, with the corresponding
probability amplitude $A[i]$.
 

\begin{thebibliography}{10}
\bibitem{FOOT} Note that the probabilities, constructed from these amplitudes, must add up to less than one.
At each beam splitter a photon may go "the wrong way", and never reach the detector $D$.
\end{thebibliography}
\end{document}